# Development and Validation of an Artificial Neural Network for the Recognition of Custom Dataset with YOLOv4

Parsa Veysi*, Mohsen Adeli**, and Nayerosadat Peirov Naziri***

* Department of Mechanical Engineering, Ruhr University of Bochum, Germany

*Abstract*—The expansion of application possibilities, which are used by more and more people, increases the performance of hardware systems and the further development of cloud systems in the field of data processing for big data. This results in countless unexplored areas of application for deep learning applications, including somewhere research is already very advanced. The field of computer vision for automated object recognition has increased many times over using deep learning algorithms. The possible uses of image processing using deep learning range from the recognition of elements with varying shapes and sizes, to the classification of complex elements and structures, recognition with varying backgrounds and changing lighting conditions, to text recognition. The advantages of using deep learning are the robust setup and high performance when recognizing complex elements. The aim of this work is the development of a detection system using deep learning algorithms for the automated detection of different components from an assembly assembly, which differ slightly to greatly from one another in characteristics such as geometry, size, contour, or color. The implementation is carried out using the single stage detection algorithm YOLOv4 (You Only Look Once, Version 4) to detect the components based on their characteristics. The quality of the self-developed detection system is checked using different test scenarios for the 13 components. The recording of the components in different orientations and different numbers, of individual parts or assembled groups is carried out using a Raspberry Pi microcontroller with a camera. In particular, the correct object recognition and the confidence value of the recognition of objects based on different positioning and views as well as the different composition of the objects, whether individually or as a group, with different distances to each other are checked and assessed. Environmental factors that can impair the quality of the detection system as disruptive factors are also to be examined. Based on the evaluation of the test scenarios, the embedding of such an application for automated object recognition for industrial use is assessed. The object recognition was very positive in all test scenarios, regardless of the orientation and the number of objects. Even with compactly assembled objects that were very close together, all objects could be recognized correctly, and the confidence value was very high at 97 to 100 percent. Likewise, all objects could be captured and labeled with the object bounding box. Shading of the objects due to the lighting conditions also did not seem to affect the quality of the results. The developed detection system is therefore suitable for real-time detection with two-dimensional image evaluation of components with different characteristics. For research purposes, an extension of the developed detection system regarding a three-dimensional image analysis would be conceivable. A conversion would take place over several cameras with different positioning and different views.

## I. Introduction

The use of Deep Learning algorithms for industrial applications is rapidly increasing. A trend analysis shows an exponential increase since the year 2014 [1]. The background is the equally exponential growth of data as more and more people are using smartphones, social networks (e.g., Facebook, Instagram) or streaming services (e.g., Netflix, Amazon Prime). Deep Learning algorithms have better performance than conventional learning algorithms for processing the performance for processing the generated data. The application possibilities are enormous. The application areas, e.g., image recognition, localization, or object recognition, which are of interest for the treatment of the topic in this work, extend to all industrial applications [2] with increasing further development of new, more robust Deep learning algorithms, and their field of application.

Deep learning algorithms are already used in numerous industrial applications [3]. Applications range from the automotive industry, e.g., in automated driving [4], to medical technology, e.g., for the analysis of medical images [5].

In this work, unsorted components from an assembly group are considered, which mainly differ from each other in geometry, contours, or color. Dynamic feature discrimination provides the automated detection system with sufficient clues to identify the detected components. The detection of unsorted components using deep learning algorithms is quite new and new applications for such applications in this field are constantly emerging [6]. Besides the technical benefits of an automated application, it could play an important role in terms of resource scarcity and thus sustainability. Instead of disposing of unnecessary, damaged, or redundant components, they could be detected and automatically sorted out with an automated detection system that uses deep learning algorithms for object recognition.

This considers two important factors: Firstly, as a result of the scarcity of resources, the purchase prices of goods are increasingly higher. On the other hand, the waste of resources is avoided, and work and production are more friendly regarding the environment. In this paper, an application for automated object recognition of unsorted components within industrial work processes, which works with deep learning algorithms, is developed, evaluated, and analyzed.

II. A Brief Overview of Artificial Neural Network

*A. Artificial Neural Networks*

Artificial Neural Networks (ANN) are a technical approach to replicate the biological neural network of a human brain to recognize complex structures and patterns from large amounts of data and to solve problems such as image processing (object recognition) in the same way [7]. The human brain consists of approximately 100 billion nerve cells (neurons). Figure 1 shows a highly simplified structure of the natural nerve cell of a human being, in which the nuclei are strongly interconnected by the axons and the dendrites. Received information reaches the nerve cells via the dendrites, which is subsequently processed in the nucleus and transmitted to the next nerve cell via the axon. This information is transmitted in the human brain in the form of chemical and electrical signals. Learning takes place through the formation of new connections between the nerve cells.

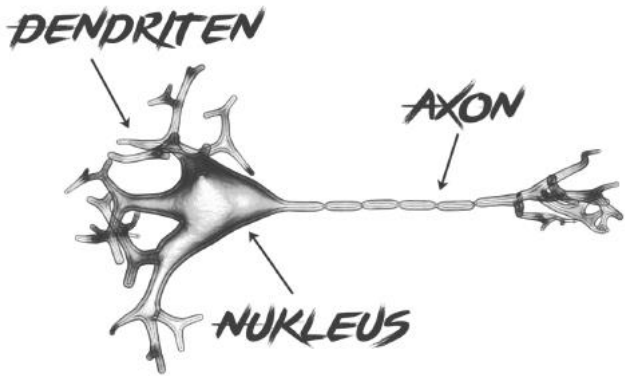

*Figure 1: Greatly simplified structure of the natural human cell [29]*

Research into ANN began in the 1940s. Its breakthrough can be attributed to the introduction of the perceptron, which is still the foundation of ANN today, in 1957 by Frank Rosenblatt [8]. Figure 2 shows the simplified structure of the artificial neuron, which consists of three distinct layers. These are interconnected by nodes, also called predominantly neurons. The first layer is the input layer. In it, the input neurons process the incoming information (data) and pass it on weighted to the next, the hidden layer. The hidden layer is positioned between the input and output layers. It can be composed of any number of layers of neurons. The received information (data) is again weighted and passed through the neurons to the output layer. This process is also called a black box because, unlike the input and output layers, the incoming and outgoing information (data) is not visible. The third and final layer is the output layer. In it are the output neurons, which contain the resulting decision as an information flow.

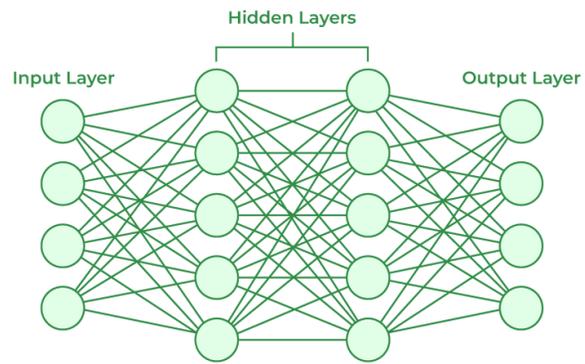

*Figure 2: Construction of a simple artificial neural network with only one hidden layer [29]*

For solving complex problems, which are usually nonlinear, neurons are linked in networks [9]. In addition to the different numbers of neurons, these links serve to distinguish the direction of signal processing. In general, the neural network consists of a set of nodes, called neurons, and directed edges. The directed edges are used to indicate the direction of the excitation signals and the order of neuron activation of the network graph. The weighted edges serve as transmission weights of the stimuli for potential regulation. The so-called recurrence networks arise as a result of the excitations, which are also considered as stimuli, but are not discussed further here because they are irrelevant for this study.

*B. Deep Learning*

Deep learning can be considered part of machine learning. It enables computer systems to transform simpler concepts into more abstract and complex ones [10]. Deep learning models, colloquially known as deep neural networks, employ multiple hidden layers to exploit the unknown structure in the input distribution and discover composite representations. Multilayer deep neural networks have existed since the 1980s, but in recent years, advances in powerful computation and the availability of larger datasets have increased the popularity of deep neural networks [37]. With the introduction of graphics processing units, training deep neural networks with greater efficiency has become a reality. Unlike classical pattern recognition systems, deep learning minimizes the need for hand-crafted machine learning solutions many times over.

*C. Convolutional Neural Network*

Convolutional Neural Networks (CNN) are the most impressive form of artificial neural networks. CNNs are the key component of Deep Learning and are mainly used to solve the difficult image-driven pattern recognition tasks [11]. The basic idea of CNN is inspired by a feature of animal visual cortex called receptive field. Receptive fields act as detectors that are sensitive to certain types of stimuli, such as edges. In image processing, the same visible effects can be produced by convolutional filtering. A typical CNN is constructed by repeating three basic types of layers: Convolutional layers, pooling layers, and fully connected layers. A deep neural network stacks many these layers to perform pattern recognition and detection tasks. A simplified CNN architecture for MNIST [12], handwritten digit image classification, is shown in Figure 3.

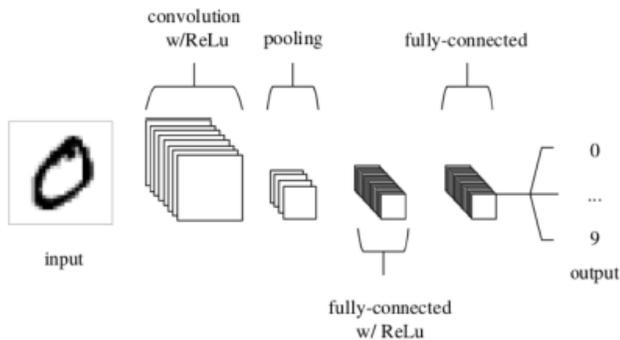

*Figure 3: A CNN architecture of five layers [29]*

Although CNNs have attracted much attention in recent years, their history dates to the 1980s. The first supervised learning algorithm with a gradient descent, was used by Rumelhart et al. in 1986 and later by LeCun et al. for handwritten digit recognition. However, this algorithm suffered from several performance problems. One of the problems related to poor built-in invariance, as it could not handle variability in handwriting patterns in translations or distortions [35]. This led to the creation of a new model with stronger shift invariance that responded to hierarchies of local features. This new network was called the Convolutional Neural Network (CNN). CNN were abandoned after a short time due to lack of computational power and replaced by Support Vector Machines (SVM) [36]. However, the development of powerful GPU in the last decade has rekindled hope in the perceptual capabilities of CNN. Today, CNNs are used as a standard approach to solve many computing problems. The operation of a CNN can be described as follows: A CNN consists of several layers of operations, each with its own function. The CNN takes an image as input and feeds it to the first layer. Feature information is passed through the hidden layers. For each layer, the activation functions perform element-by-element activation of the output generated by the previous layer. The output of the last layer is then compared to the target output. This process is called a forward pass.

### C. Computer Vision

The year 1966 represents the birth of so-called computer vision [13]. This is the processing and understanding of digital images and videos by machines. Various filtering techniques, such as object recognition, are used to make decisions for different types of questions and problems. Inspired by the idea that machines, as the spitting image of the human visual system, can perform the same activity, this idea led Wiesel and Hubel to research the information processing of the visual system of cats [14]. In doing so, they investigated their visual perception in the presence of different images. They concluded that the cat brain responded primarily to the change of images rather than to their content. The first visual processing by machines started by edge detection of different structures. For this, Torsten Wiesel and David Hubel received the Nobel Prize in Medicine in 1981.

In the 1960s, the first two laboratories in the field of artificial intelligence were opened at the American universities Massachusetts Institute of Technology and Stanford University [15]. In 2001, Paul Viola and Michael Jones introduced the Viola Jones Method, a real-time face recognition algorithm used by Fujifilm for the first digital cameras with face recognition [16].

### III. OBJECT DETECTOR ARCHITECTURE OF YOLOv4 ALGORITHM

You Only Look Once Version 4 (YOLOv4) is a real-time CNN for object recognition. The network predicts bounding boxes and class probabilities from images in an evaluation. The real-time aspects arise from the fact that detection is framed as a regression problem. As a result, no complex pipeline system is required, so detections are predicted by simply running the network on an image. There are five versions of YOLO in total, but only the first four [17] [18] are supported by a scientific paper at the time of writing. Therefore, the most recent scientifically supported version will be used, namely YOLOv4. YOLOv4 was released on April 23, 2020. YOLOv4 comes in a small version that focuses on systems with limited resources. This model applies the same techniques as in YOLOv3 but has fewer convolutional layers. Figure 4 shows the prediction on an image using the YOLOv4 algorithm.

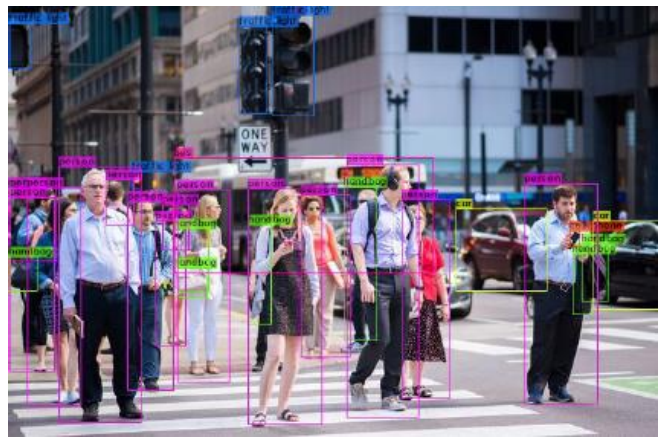

*Figure 4: Object detection with YOLO algorithm [30]*

YOLOv4, released in April 2020, changed developers because the previous developers had stopped their efforts in computer vision research. They were concerned about how the technology was being used for military applications and worried that privacy concerns would have societal implications. This version mainly combines state-of-the-art methods to improve YOLOv3.

### A. Input and Output

YOLOv4 processes input images with a resolution of N x N pixels and three channels. The pixel resolution N must be a multiple of 32. The authors of YOLOv4 used three different resolutions for their experiments: N = 416, N = 512, and N = 608. A higher resolution input image not only leads to higher accuracy, but also to higher training and inference times. Most of the publicly available pre-trained YOLOv4 models are trained with resolution N = 512. The examples shown here use the resolution N = 416.

The network predicts objects at three different scales. This means that at the feature extraction point of the network, feature maps are extracted at three different levels. Since the feature extraction part consists mainly of convolutions, input images become smaller and smaller as one goes deeper into the network. Thus, extracting feature maps at different points preserves high, medium, and small features. This is useful for detecting objects of different sizes, e.g., cars are relatively large, so detection with small features (lower resolution) is favorable. On the other hand, the detection of small objects like traffic lights can be done by the high feature maps (high resolution). Figure 5 illustrates the idea of extracting features at different levels. The size of an output stage $N^*$ at each stage i is defined as:

$$N_1 = \frac{N_{in}}{8}, \quad N_1 = \frac{N_{in}}{16}, \quad N_3 = \frac{N_{in}}{32}$$

Each output pixel in the output feature map, now called a grid cell, is a 1D tensor that predicts the position and class of an object. The 1DTensor consists of four predicted coordinates for each bounding box $t2, t3, t\&, t4$, and an object confidence value $p$ (confidence value). Each 1D tensor also predicts $C$-conditional class probabilities. This results in the tensor containing the following predicted tuple: [$(t2, t3, t\&, t4), p5, (C1, C2, ..., Cn)$]. Since the output of stage i consists of $N^*$ grids, there is a 3D tensor with $N^* \times N^*$ 1D tensors. These 3D tensors are referred to as boxes. Each level predicts three boxes, see Figure 5.

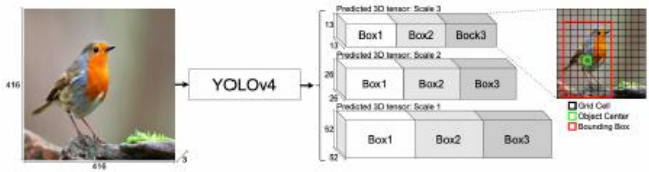

Figure 5: Overview of the YOLOv4 process [31]

The center grid cell of the object's ground truth bounding box is responsible for predicting the object. The object value of this grid cell is one and zero for others.

### B. Bounding Box Prediction

Each bounding box in the original YOLO consists of four predictions: $x, y, w, h$. The center of a box was represented by $(x, y)$-coordinates relative to the boundaries of the grid cell. The width $w$ and height $h$ are predicted relative to the whole image. This approach changed in the second version of YOLO by using bounding box priors (anchors) and predicted offsets instead of coordinates. Predicting offsets instead of coordinates simplified the problem and made learning easier for the network.

The anchors are initialized with two prior anchor dimensions: Width $p\&$ and Height $p4$. The network uses these bounding box priors to predict the Height $t4$, Width $t\&$, and Center coordinates $(t2, t3)$. Figure 6 provides a graphical representation of the anchor-based learning problem. The following equations transform the predictions to obtain bounding boxes:

$$b_x = \sigma(t_x) + c_x$$
$$b_y = \sigma(t_y) + c_y$$
$$b_x = p_w \cdot e^{t_w}$$
$$b_y = p_h \cdot e^{t_h}$$

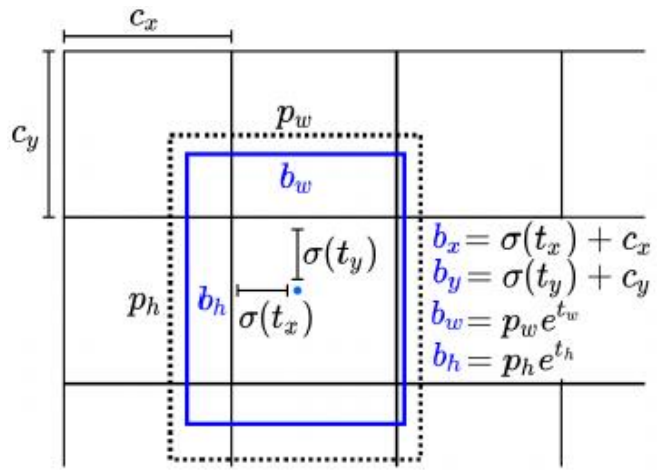

Figure 6: Anchor box [32]

### C. Architecture

The YOLOv4 architecture consists of three parts, a backbone for extracting features, a neck used to collect feature maps from different phases, and a head that predicts classes and bounding boxes of objects. Figure 7 shows the architecture. In the next sections, each part (Backbone, Neck, and Head) is described in more detail separately.

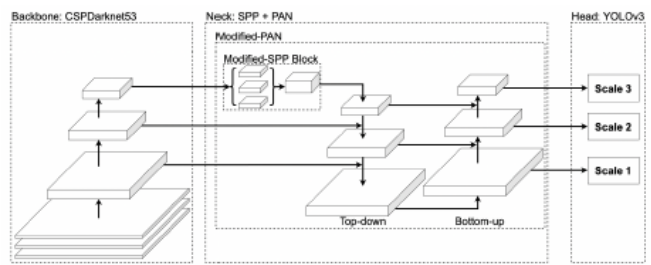

Figure 7: Overview of the YOLOv4 architecture [22]

### D. Backbone

Extracting features from the input images is the first step of the network. For this step, YOLOv4 modifies the Darknet53 CNN as used in YOLOv3. The Darknet53 network uses consecutive 3x3 and 1x1 convolutional layers and skips connections known as residual connections [19]. Modifying Darknet53 by implementing cross-stage partial networks (CSP) [20] leads to using the network of YOLOv4, CSPDarknet53. This network consists of five CSP blocks, which in turn use $n$ residual blocks. Before each CSP block, the input feature map is down sampled by a convolutional layer. Feature maps are extracted in three different phases, after the third, fourth, and fifth CSP blocks. A complete overview of the CSPDarknet53 is shown in Figure 8. The backbone is trained separately from the entire YOLOv4 network on the ImageNet dataset. Before training, an average pooling layer, a fully connected layer and a non-linearity layer (Softmax) are added.

### E. CSP block

A cross-stage partial block (CSP), blue in Figure 8, splits the data channels into two parts $x = [x6, x66]$ and then merges $x66$ with the original computation at $x6$. This splitting and merging of data has several advantages. First, the gradient path is doubled by the split-and-merge strategy. In addition, there is a reduction in the amount of memory traffic since only a portion of the original computation is processed. The authors of YOLOv4 added additional convolutional layers to each branch and finally performed convolution on the concatenated feature map. These so-called transition layers maximize the difference in gradient combination.

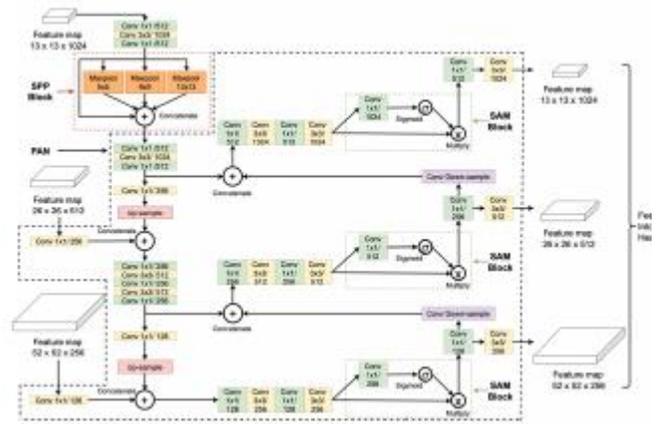

Figure 9: YOLOv4 Neck [22]

### H. Head

YOLOv4 employs the same header as in YOLOv3. Each feature map received from the neck passes through a fully connected layer implemented as a $N^* \ x \ N^* \ x \ F$- Convolutional Layer with 1x1 filters, where $F = 3 \ - \ (4 + 1 + C)$. Output F represents the 3D tensor with three boxes, $N^* \ x \ N^*$ 1D tensors, consisting of four bounding box coordinates, an objectness value, and C-conditional class probabilities. Figure 10 shows the head of YOLOv4.

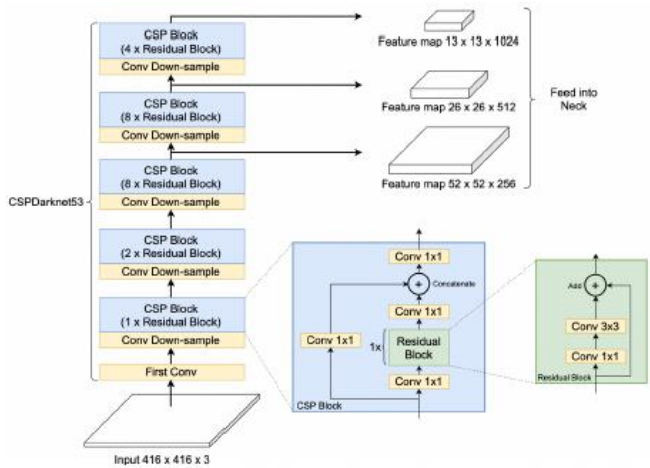

Figure 8: YOLOv4 Backbone [22]

### F. Residual Block

Residual blocks offer a solution to vanishing or exploding gradients in deep networks. Networks do not perform better by simply stacking more layers, as the inventors of the residual block have shown. So, they experimented with skip connections that perform identity mapping on their outputs. Skipping a connection is mathematically defined as $y = F(x) + x$, where $x$ is the input (identity), $y$ is the output, and $F()$ is the feature mapping. This identity mapping technique does not add additional parameters or computational complexity but increases the accuracy of deep networks. The green block in Figure 8 represents a residual block. The feature mapping function $F()$ performs the original darknet 3x3 and -1x1 convolution. The input is then copied to a separate branch, and both are added at the end.

### G. Neck

The backbone is followed by the neck. Its goal is to enrich information fed from the different stages by the Backbone and passed to the Head. The Neck modifies and combines three different state-of-the-art methods to realize this: a Path Aggregation Network (PANet) [21], an SPP block, and three SAM blocks. Figure 9 provides a graphical overview of the Neck. Each block is discussed separately in this section.

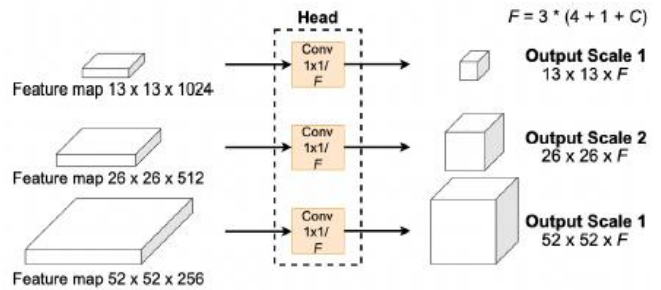

Figure 10: YOLOv4 Head [22]

### I. Mish Activation Function

Mish [34] has been proposed to improve the performance and address the shortcomings of ReLU, just like Leaky ReLu. The researchers [22] of Mish found that Mish achieves or even improves the performance of neural networks compared to ReLu and Leaky ReLu in various computer vision tasks. Figure 11 shows the performance of Leaky ReLU and Mish in comparison. The equation in figure 11 defines the Mish activation function mathematically.

$$f(x) = x \cdot tanh(softplus(x)) = x \cdot tanh(ln(1 - e^x))$$

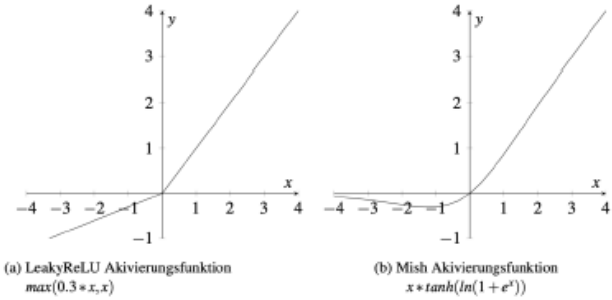

Figure 11: LeakyReLU and Mish activation functions compared [22]

*J. Processing Output*

The network outputs predictions on three scales, each with Ni x Ni grids. By summing all the grids, the total number of objects that can be detected is determined. For example, using a 416 x 416 input image, a total of 3549 predictions are calculated across the three scales. Filtering these predictions, keeping only relevant predictions, is an important post-processing step. One technique that is often described in the literature is the nonmax suppression algorithm. This algorithm consists of two steps. First, predictions with an objectivity score below a certain threshold are removed, then bounding frames with an IoU higher than or equal to a certain threshold are discarded relative to a bounding frame with a higher objectivity score. Figure 12 represents the nonmax suppression algorithm in pseudocode.

```
Input  : B = {b_1,..,b_n}, P = {p_1,...,p_n}, λ_P, λ_IoU
         B is a list of bounding boxes
         P contains corresponding objectness scores
         λ_P defines the objectness threshold
         λ_IoU is the IoU threshold
Output: B_r = {}, P_r = {}
         B_r is a list of non-max supressed boxes
         P_r contains corresponding objectness scores
begin
    B_r ← {}
    P_r ← {}
    /* Discard all boxes with objectness under the threshold */
    for b_i in B do
        if b_i < λ_P then
            B ← B - b_i
            P ← P - p_i
        end
    end
    /* Discard boxes with high IoU relative to a box with a higher objectness score */
    while B ≠ empty do
        P_max ← max(P)
        B_max ← b_{P_max}
        B_r ← B_r + B_max
        P_r ← P_r + P_max
        B ← B - B_max
        P ← P - P_max
        for b_i in B do
            if IoU(B_max, b_i) ≥ λ_IoU then
                B ← B - b_i
                P ← P - p_i
            end
        end
    end
    return B_r, P_r
end
```

Figure 12: Nonmax suppression algorithm in pseudo code

## IV. STATE OF THE ART

In this paper, the current state of the art regarding machine learning using deep learning algorithms is addressed. Models and data sets used in the field of image processing and object recognition are taken up.

*A. Transfer Learning*

Transfer learning is the use of a pre-trained model that is transferred to another task [23]. Either the complete model is transferred or only some initial layers. This does not mean that only the architecture of the model is transferred, but the weights of the pre-trained model are transferred. There are two common approaches to transfer learning. In the first approach, the pre-trained model is used only as a feature extractor [24] [25]. This means that it is used to generate features up to a certain layer. These features could then be used as input in another machine learning algorithm to train a different task. It is also possible to build some more layers on top of the transferred layers, which are then learned, but the transferred layers remain frozen. So, the task of the pre-trained model is only to extract the features that are then used to train the task. The second approach is fine tuning. The pre-trained model or only a part of the model is transferred, and more layers are added. However, now the weights of the transferred layers are also trained. It is possible to keep some lower layers of the transferred model frozen. The benefit of this transferred learning is that the pre-trained models are usually trained with large data sets containing millions of images. This has the advantage that the model has learned useful features in the lower layers. These features could be suitable for similar tasks. A positive side effect is that training is faster when the pre-trained model has been applied to similar tasks because the weights are well initialized. In addition, it can help improve performance on small datasets because the weights are well initialized to detect common features in images. Yosinskiet et al. [26] showed that transfer learning and fine-tuning improved the generalization of their tested datasets.

*B. Microsoft COCO dataset*

The Microsoft Common Objects in Context (COCO) dataset [27] is a popular collection of images of everyday scenes. It is used to extend the state of the art in terms of image processing. In total, the COCO dataset contains 328,000 images with 2.5 million labels of objects assigned to 91 defined classes [27]. The classes are chosen to be easily and uniquely identifiable. They are easily identifiable objects such as "bird", "cat", "dog" or "car".

*C. Darknet*

Darknet is an open-source neural network framework written in C and CUDA. It is easy to install and supports CPU and GPU computations. Darknet is mainly used for object recognition and has a different architecture and features compared to other deep learning frameworks. The approach Darknet uses is faster compared to the architecture of other neural networks, e.g., Faster R-CNN etc. They need to be written in C if speed is required, and that is the case with most deep NN frameworks. Tensorflow, for example, has a broader scope, but Darknet Architecture and YOLO are specialized frameworks and are at the top in terms of speed and accuracy. YOLO can run on the CPU, but it results in 500 times more speed on the GPU from using CUDA and cuDNN.

## V. TEST EXECUTION

The goal is to create a deep learning model in which the selected algorithms recognize all object classes using the appropriate libraries. To improve the training results in terms of accuracy, the created dataset is manipulated using a dataset expansion and more datasets are artificially created. For the visualization of the results, a self-written program is connected to the learned trained network. Finally, the entire network including the algorithm is analyzed using the results from the test scenarios.

### A. Generating The Training Data

The objects differ in size and geometry, which is why the image size of approximately 608 x 608 was selected for the images of all classes. Due to the varying characteristics of the objects, a side view would not allow all of them to be detected correctly. Therefore, all images are recorded from the front view perspective, for both the training sets and the validation sets. Since distortion of the images or overlapping of the objects would lead to a significant degradation of detection, making it difficult to unambiguously match the objects, these scenarios are excluded for network assessment. Changing the distance from the camera lens to the object may also result in a loss and be reflected in the number of false detections. A predefined distance between the plane on which the objects to be detected are located and the camera was deliberately chosen for the best results.

The recordings should represent a realistic image, similar to an industrial work process. For this reason, all objects were photographed in a windowless room with the support of an LED ring light. One advantage is the non-varying disturbing factors, e.g., avoiding the creation of different shadow conditions due to a differing incidence of light. For the training set, all objects were recorded individually, in different positions and inside view with the camera. By using many non-redundant images for each class, the set can be trained effectively.

After creating the training data, a class for each component is created for these images using LabelImg software, a graphical image annotation tool for labeling object bounding boxes in images. Based on the object bounding box, the weights of the loss function are adjusted. An object bounding box is obtained from the four coordinates $x_{min}$, $x_{max}$, $y_{min}$ and $y_{max}$, which are used to define the frame of an object detection. For correct detection per class (objects), at least 300 data (images) are necessary for the neural network to have enough intuition to correctly recognize the corresponding classes (literature). The network is given the training data to learn the specified classes. Through the different positioning and views for each object, an understanding is impressed on the neural network for correct recognition of the objects based on their features. The number of images should be enough for the classification of all classes from the training set to finally recognize in the validation set all learned classes, which occur arbitrarily in different numbers assembled in groups, in different positions and views from the records.

### B. Data Expansion of The Training Data

Data augmentation of training data with artificially generated data sets is a popular technique in computer vision when data sets are insufficient. Different techniques, e.g., image rotation, contrast change, image size change or zooming within a recording, are used to manipulate the data set. In the previous sub-section, the conditions to guarantee a realistic image, were described. Also, the predefined distance was mentioned. Therefore, the techniques of contrast change, image resizing and zooming within the recording are excluded.

To increase the robustness of the algorithm, the technique of image rotation was applied. Here, each object is recorded once from each view or orientation and each recording is rotated clockwise. In this process, any number of recordings of an object can be made within the 360°. This technique is extended by the tilt technique, which allows recordings to be flipped about the horizontal and vertical axes. Figure 13 shows the technique mentioned using an object from the training set as an example.

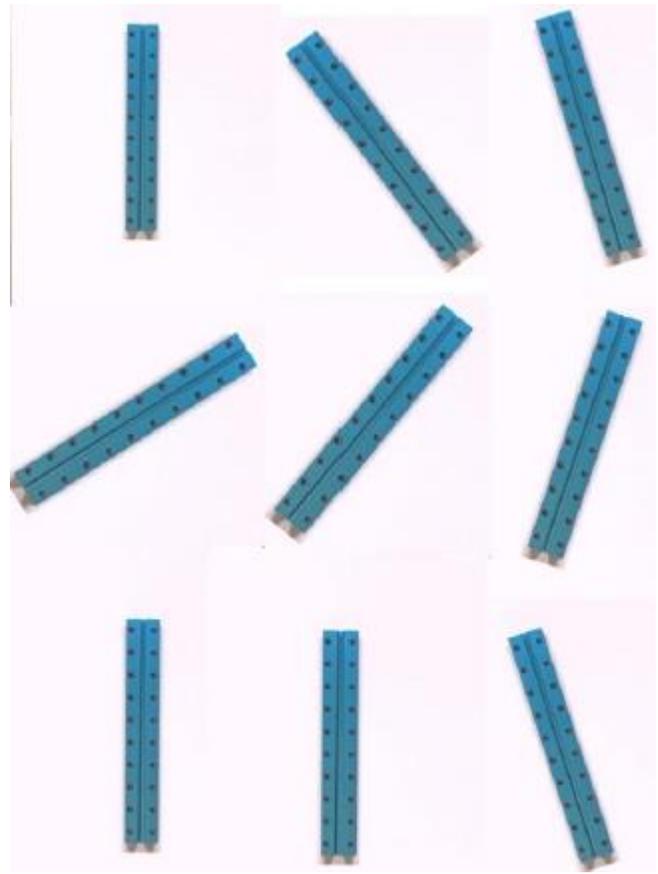

*Figure 13: Data extension of a component from the training set*

Since the recorded objects differ from each other based on their characteristics, different numbers of records for each object are generated as a result of the different orientations. For each object, at least 300 images were generated by the presented technique of data augmentation. All images of the 13 recorded components were transferred to Darknet for integration in the training process using the computer vision library OpenCV.

*C. Processing The Training Data*

First, all images recorded with the camera must be labeled. For the objects in this work, the object bounding boxes must be manually created to precisely encompass the contour of the objects. For the training set, all objects are labeled once from each view with an object bounding box. Figure 14 shows a labeled object bounding box of an object from the training set.

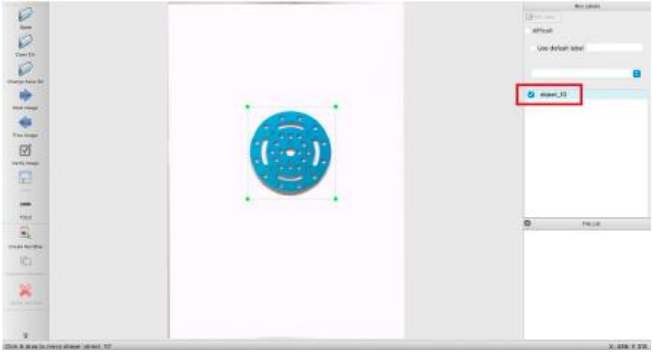

*Figure 14: Capture and label a component in LabelImg*

The *LabelImg* software creates a text file for each labeled object with object bounding box as an output of this labeling. In it, the coordinates ($x_{min}$, $x_{max}$, $y_{min}$ and $y_{max}$) of all created object bounding boxes are given for all objects on the image with the corresponding class membership (labeling). The bottom Figure 15 shows all the information of the label for the object from Figure 14.

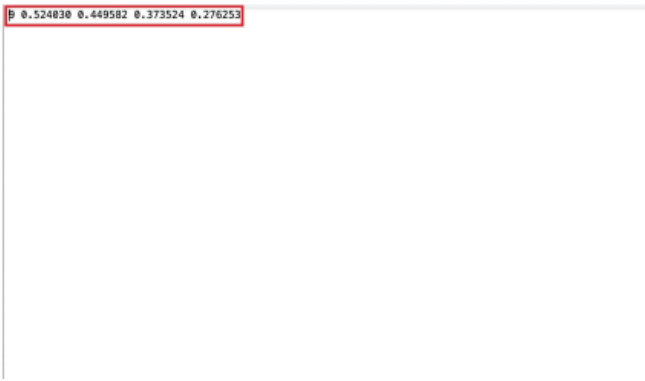

*Figure 15: Information of the label after labeling in LabelImg*

The first column represents the object name or class membership and the remaining columns represent the object bounding boxes ($x_{min}$, $x_{max}$, $y_{min}$ and $y_{max}$). The format of the created text file is called YOLO format. It is very popular in object recognition. In a created text *file,* all classes are automatically recorded. Subsequently, for each view of an object, the record expansion and labeling automation (in YOLO format) is created using a self-written program. For all collected information in the form of a text file, it is necessary to convert it into a CSV file. This summation of all information must be passed to Darknet for readout in the form of binary records. In the validation set, all 13 classes with different numbers of objects can be detected in the maximum case. The library used in the Darknet environment using the YOLOv4 algorithm is OpenCV, which contains the basic computer vision algorithms for image and video processing.

*D. Training The Network*

The goal of the training process is to match the weights from the pre-trained network [28]. Using the pre-trained COCO dataset, the network is able to detect patterns from already learned contexts. One advantage of using a pre-trained network is the detection of basic patterns that have already been learned. The weights in the network are tuned to detect up to 1000 different classes. Thus, when the weights are adjusted to detect their own objects, better convergence is achieved. However, the configuration of a network affects the speed of the training process and thus the results of the training run. The training process was computed online on Google Colab version Pro+ on the graphics card (GPU) NVIDIA Tesla T4 with 52 GB RAM, which is characterized by a faster processing time with a large amount of data.

*E. Definition of The Industrial Test Scenarios*

Based on the test scenarios, an in-depth analysis of the YOLOv4 algorithm for the deep learning model is aimed at. With these test scenarios, the goodness of the model for an application in the field of industrial automation process for different unsorted components, which can occur in different quantities, is tested. After evaluating the results - i.e., whether the network correctly recognizes all objects within randomly composed groups - the use of the network as a technical application for industrial use will be discussed. To check whether the model meets the quality criteria, the trained network is tested using the evaluation set. For this purpose, different scenarios are created that consist of different classes. Within these scenarios, a class can occur multiple times. In general, the network adapts to the data. If the test accuracy is only minimally worse than the training accuracy, the model may probably be a bit more complex. If the test accuracy is much worse than the training accuracy, the model will fit the training data too closely and overfitting will occur because the model appears to be too complex. If the accuracy is poor on both runs, more data will be needed. With the first scenario, the network is tested regarding the recognition of the individual classes. This is to prove that all classes are recognized unambiguously. For this purpose, all 13 objects are examined individually in different positions and views. This also corresponds to the training scenario with which the network was trained to recognize all objects correctly. One difference to the training scenario is the varying number of objects of a class within the first scenario.

*F. Scenario 1: Single Classes in Different Positions and Views As Single Object*

In the first scenario, 50 images are used for all classes as individual objects. Each object from the respective class is recorded in different positions and views. Since the objects differ in their geometry, resulting in different positioning and especially views, a different number of recordings per class is created. Each object from the different classes was rotated and recorded from different views. Figure 16 shows some of the different positions from different views of the objects from the dataset.

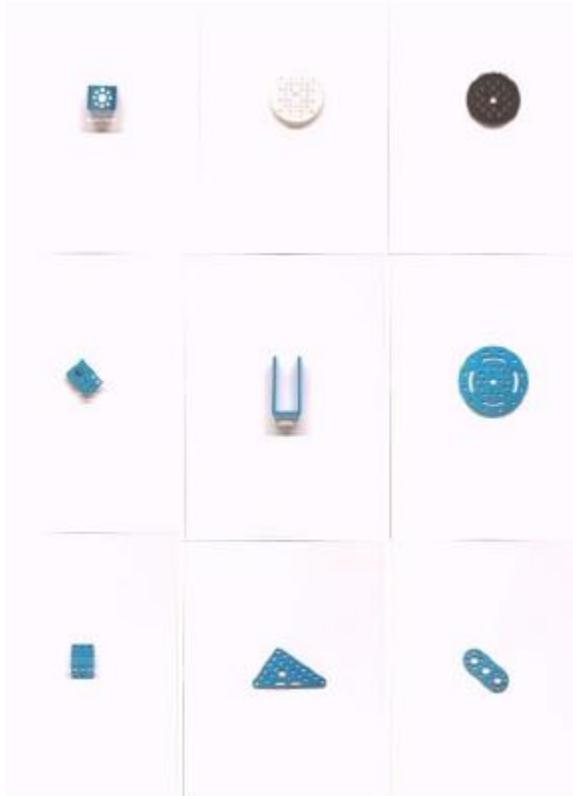

*Figure 16: Objects from the data set*

The goal is to use different orientations of all objects to ensure that the network can detect and correctly recognize all classes. The only interfering factor that cannot be completely avoided is the shading of objects due to lighting conditions. Since a shadow is not a feature of an object, the detection of an object must separate itself from it. By using a two-stage control factor, the creation of false object bounding boxes should be avoided. In the first stage of the control factor, it is ensured that the detected object is assigned to the class with the highest match. In the second stage, the object bounding box and label for the detected object are created for the highest matching class only if the network has detected the object with a predetermined level of detection. Based on the output of this two-stage control, a judgment can be made about the quality of the network, i.e., the correct object detection, for all three scenarios.

### G. Scenario 2: Multiple Classes in Different Positions and Views As A Group

With the second scenario, the complexity from the first scenario is extended. The network must be able to distinguish objects from each other despite different possible combinations within a recording. In addition to different orientations, these can be recorded in varying numbers. The distance between the objects is deliberately chosen arbitrarily. In some cases, they can be very close to each other. Based on the detection of multiple objects within a recording, conclusions can be drawn about the training process, especially about how well the network has learned to recognize relationships between individual classes. Figure 17 shows different groupings of objects from different classes in different orientations with varying numbers of pieces.

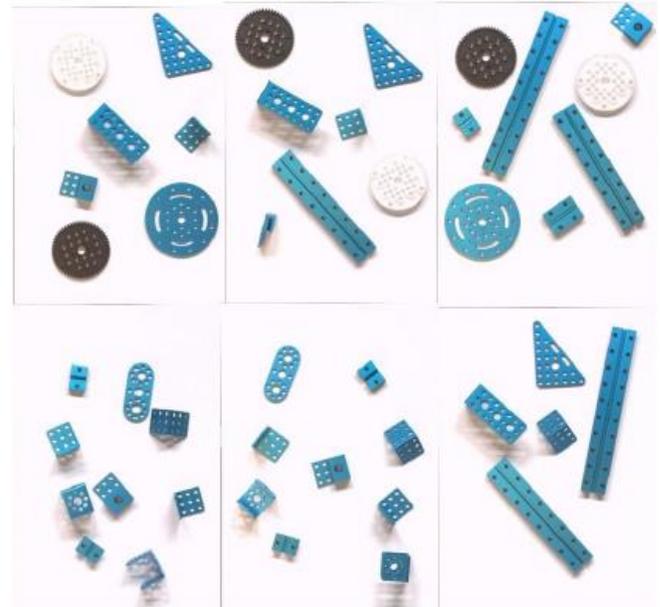

*Figure 17: Different groupings of objects from different classes in different orientations with varying numbers of pieces*

The second scenario is intended to test the network's ability to differentiate. Specifically, it should be determined at which points, i.e., for which classes, the network has problems with differentiation. To evaluate the network, it is checked whether all objects were correctly recognized and with what accuracy this was done. Based on the accuracy of the object recognition, conclusions can be drawn as to which object orientations are more difficult for the network to recognize.

### H. Scenario 3: All Classes in Different Positions and Views As A Group

In the third and last scenario, all objects from all classes appear in all records in different orientations and views. Here, too, a varying number of objects can occur from some classes. Since the objects - as in reality - are put together arbitrarily, different distances between them can arise. Thus, a direct juxtaposition cannot be ruled out either. In this scenario, the complexity from the second scenario is significantly increased. It should be determined here whether and at which points the network has problems with differentiation. Figure 18 shows different compositions of all classes from different perspectives and orientations.

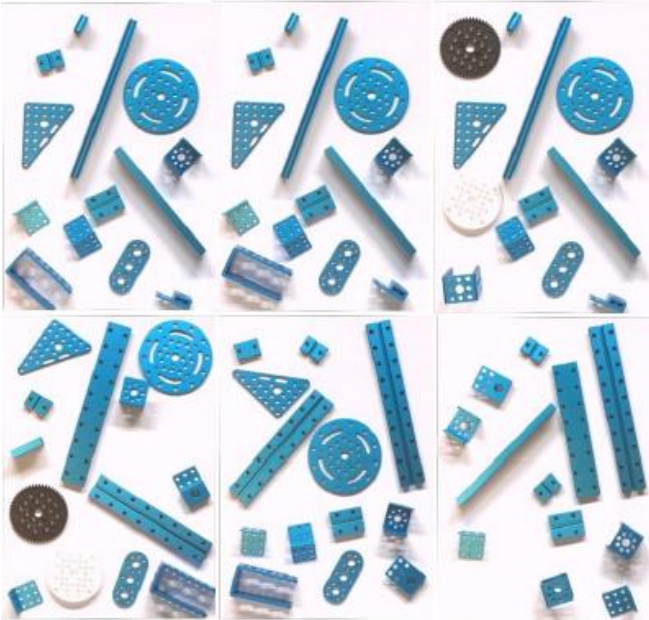

*Figure 18: Different compositions of all classes from different perspectives and orientations*

## VI. EVALUATION

In this section, the results of the set up deep learning model are presented and analyzed in depth. First, the results from the training are evaluated. Subsequently, the set-up network is evaluated using the evaluation set. For this purpose, in particular the selected approach for the set-up training condition is evaluated using the YOLOv4 algorithm. For this evaluation, the test scenarios are examined in detail and based on these examinations, it is argued to what extent the set-up detection system is suitable for industrial applications. The limitations and thus the weaknesses of the system with respect to the automated detection of components with different features are discussed.

### A. Study of The Developed System

YOLOv4 models use a fixed input size, usually 416 * 416 or 608 * 608 pixels. The number of the neurons in the input layer is determined by the dimensions of the image multiplied with the 3 color channels, assuming a standard RGB image. The input size (dimensions) of each image of the dataset is 608 * 608 pixels. Thus, input layer of the model has 1108992 neurons. YOLOv4, implemented in the Darknet framework, typically has 53 convolutional layers in its backbone network, which can be considered as hidden layers. Therefore, the model contains 53 hidden layers. The number of neurons in a convolutional layer can be calculated by (Height of Future Map) * (Width of Feature Map) * (Number of Filters). Applying this formula to each convolutional layer of the model and sum them up together to get the total number of neurons in the hidden layers. The total number of neurons in all hidden layers combined is 202264690 neurons. The proposed YOLOv4 model in the darknet framework employed the Mish activation function, due to enhance the network's performance and reducing instability compared to other functions like (Leaky) ReLu.

The overall training process using the detection weights from the COCO dataset is shown in Figure 19. The horizontal axis indicates the time units in epochs. One epoch corresponds to the evaluation of all included data (images). The vertical axis indicates the total loss function for each epoch traversed. The resulting graphical progression indicates the loss rate of the training set.

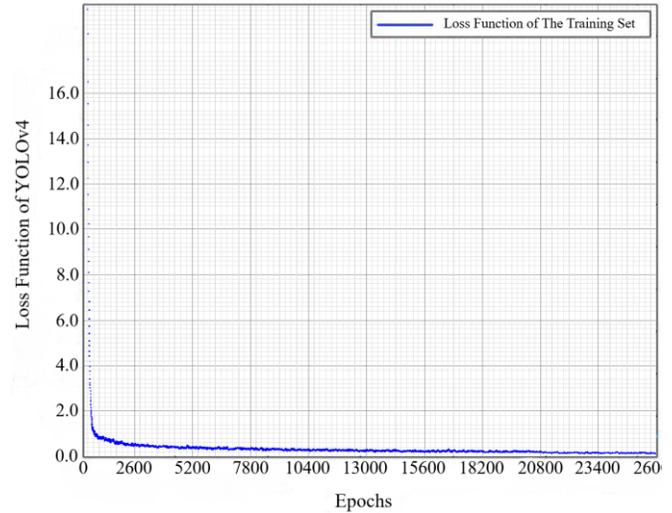

*Figure 19: Loss function of YOLOv4 training set*

Since the model does not yet know any of the new classes created, the loss rate is well over 100 percent. After the first 2600 epochs, the loss rate flattens out sharply, which is due to the pre-trained weight. This is the main reason for using a pre-trained network because the adjustments of the weights compared to a non-pre-trained network are faster. The variation in the loss rate between epochs that occurs thereafter is small for the time being. Only after approx. 21000 epochs have been run through, the loss curve flattens out again more strongly. The whole training process turns out to be stable on the trained images. In particular, the behavior of the trained model on unknown images avoids the risk of overfitting. Overfitting usually occurs when the accuracy of the training set is high, because the model has adapted too much to the training data and thus the training and test accuracies differ strongly. The test set consists of images of all classes, individually or in groups, with different positions and views. Certain interfering factors, such as background color, image contrast or manual zooming in with the pi camera, cannot be completely filtered out, which become noticeable in the second and especially the third application scenario through slight fluctuations in the result quality.

### B. Evaluation of The Industrial Test Scenarios

The industrial test scenarios from the test execution are evaluated in the following. All three scenarios are examined for susceptibility with respect to the correct detection of the objects. All confidence values of the detections are also evaluated, since in addition to the basic functionality of the object detection, the accuracy of the detection is also crucial for the evaluation of the network.

1) Based on the evaluation of the first scenario, we will prove that the network has the basic intuition to detect single classes as it has been trained to recognize and detect them in different positions and perspectives. The evaluation immediately shows that all classes were detected correctly and with a confidence value of 100 percent. Only for some classes the confidence value was 99 percent. All classes can thus be described as stable in terms of correct detection. The automatic creation of object bounding boxes also turns out to be very precise for the objects from all classes, regardless of orientation. The shading of the objects as a result of the lighting conditions, which is particularly noticeable in the side view, does not impair the quality of the results.

The result from the first scenario is very positive for the examination of all classes. All objects from the data set are correctly recognized as individual classes in different positions and views. All classes are correctly recognized based on their features, regardless of whether they are different or similar. Also, the shading does not seem to affect the correct recognition of the objects, especially since all object bounding boxes cover the objects very well. Overall, the evaluation, with a high mAP value of 0.937 (IoU = 0.50 : 0.95), is very stable.

2) In the second scenario, the complexity is increased. The evaluation checks whether the network is able to recognize different combinations of several classes in different orientations and numbers of objects. In Figure 20, different combinations of multiple classes are shown as groups. All classes are correctly recognized regardless of their orientation and number of objects. The success rate is 99 to 100 percent on average. Even in the side view of the objects, regardless of their group membership, their shading does not seem to influence the result negatively. Only with some objects do certain positioning and views seem to minimally reduce the confidence value, as this deteriorates somewhat and is at 96 percent.

Thus, even in the case of several classes with different orientations and numbers of objects, the network has no difficulty in correctly recognizing all classes with a perfect confidence value. A special feature is the distance of the objects within the randomly composed groups, which has an influence on the quality of the object recognition. If the objects are close to each other, they are all correctly recognized by the system. They are completely covered with the object bounding box, regardless of shading due to lighting conditions. For the second scenario, it can be stated that neither the number of classes nor the different positioning or number of objects on a recording negatively influences the quality of object recognition. However, if the components are too close together, the features, such as the contours, of an object cannot be clearly identified. For this, two components, where the boundary transition is difficult to identify, can be recognized as one object by the detection system. In this case, the positioning and views play a role insofar as the edges of the objects that are adjacent to each other are difficult for the system to separate. This case occurs in two of the 50 images examined. This corresponds to an error rate of 8 percent. The mAP value is 0.91 (IoU = 0.50 : 0.95), which is also very stable, as in the first scenario.

3) The third scenario represents an extension of the complexity from the second scenario. All 13 classes from the trained network are present in all recordings. They are randomly composed in different positions and views. A class can appear multiple times in a recording. The composition of the objects happens randomly and thus different distances among them result. Thus, objects from the same class or from different classes can line up. Figure 21 shows the compositions of all 13 classes in different orientations and numbers of objects.

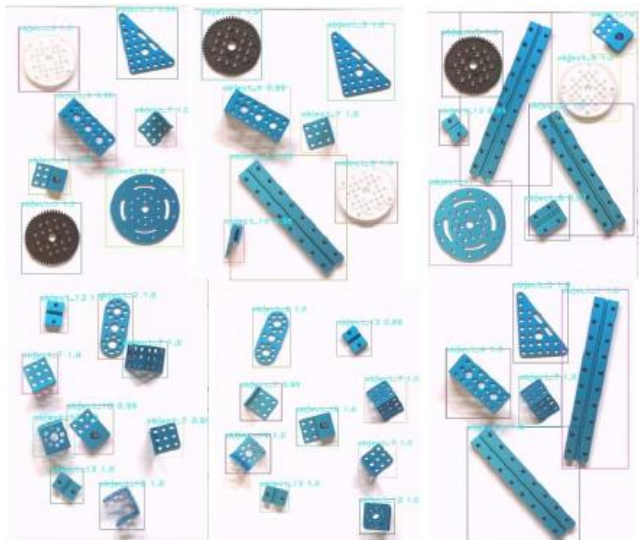

*Figure 20: Detection of different combinations of several classes as groups*

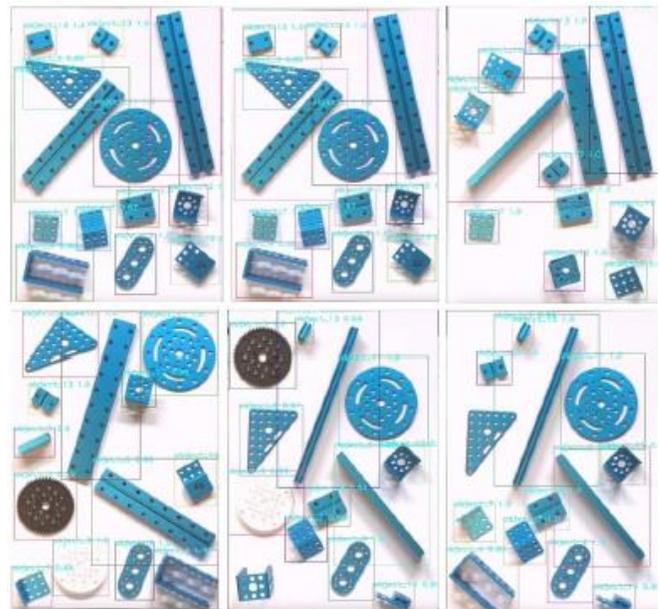

*Figure 21: Detection of all 13 classes in different orientations and numbers of objects*

Figure 21 shows that the detection system is able to recognize all classes correctly. The different shades of the components, the varying number of objects from one class or the close distance between the objects of the same class or different classes do not seem to endanger the correct object detection. The confidence value of the recognized classes is between 98 and 100 percent. Thus, it turns out stable. Even in the case of closer compositions of all objects, where they slightly touch each other, all are predominantly completely covered by the object bounding box and correctly labeled. False detections can occur in some cases, comparable to the second application scenario, with closely assembled objects. The system cannot - due to the lighting conditions and the basic color of objects - distinguish the edges, i.e., the contours, of the objects to be detected and this leads to false detections. The third scenario provides the same results as the second scenario despite the increase in complexity. The network is able to encapsulate interfering factors, e.g., shading due to light conditions when there is little contact between the objects, from the object to be detected. The mAP value is with 0.89 (IoU = 0.50 : 0.95) in relative comparison to the other scenarios, the weakest.

## VII. Evaluation of The Network on The Basis of Industrial Goods Criteria

Based on the evaluation from the training and the test scenarios, a judgment can be made about the application of automatic object recognition as an application for the industrial use of unsorted components. The applicability of the deep learning model is evaluated based on different criteria. Reliability, accuracy, and efficiency play an essential role for the application of such an application for industrial use.

### A. Reliability

One of the most important criteria for a possible application of a deep learning model for industrial use is reliability. How important it is becoming apparent even with a minimal error rate. As a result of a small deviation, e.g., in the case of an incorrect object recognition, an incorrect inventory occurs, which can lead to an impairment of the industrial work regarding the planning of manufacturing and assemblies. This in turn is not only time-consuming, but above all costly.

To avoid unnecessary interruptions in the work process, a low error rate is aimed for. If the error rate were to exceed 3 percent, which may well occur in the case of shape-like objects, e.g., beam elements, or could result from a crowded composition of objects touching each other, the detection system would be unsuitable for an industrial application. One way to keep the error rate below 3 percent is through the training process. Specifically, when the training set is created, each object view should be sufficiently signed off to provide the network with enough intuition during the training process to recognize all classes based on their features, regardless of their orientation. One factor is the small difference that can lead to false detections. This is especially true for objects with nearly identical features and likewise for those that have many positions and views due to their geometry.

A correct object recognition of all classes is not yet decisive for the use in the industrial working process, but the repetition accuracy is decisive here. For the automation of a work process, the results must be reproducible. This was confirmed by the different test scenarios 1, 2 and 3. In case of repeated occurrence of a class within a recording, i.e., if a class is present several times, this does not affect the quality of the detection and the confidence value. In the case of multiple contiguous objects, all object bounding boxes predominantly turn out to be very stable, with rare exceptions, especially for test scenario 3, where all classes are present with different orientations and numbers of objects.

### B. Accuracy

The accuracy of a detection system is primarily relevant for the correct sorting of the detected objects. Conceivable is the detection and subsequent automated sorting, e.g., of unsorted components from a component group. To avoid errors in the automated sorting and thus in the production process, the trained network should have sufficient intuition to fully detect all objects from different positioning and views. For this, the object bounding boxes for each detected object play an important role, as each object should be encompassed as accurately as possible. In all three scenarios, all objects are predominantly detected correctly with an accuracy of between 98 and 100 percent, regardless of their grouping and number of objects. In the case of very close distances, false detection can occur, or the object is not detected by the object bounding box. However, the angular object bounding frames do not adjust to the orientation of the detected object. The object bounding boxes are slightly larger for some views, e.g., the side views of some objects. However, this is irrelevant for automated sorting of components within the inventory, especially since the detection system is designed for correct sorting and counting of components.

The correct labeling, i.e., the labeling of the detected object, is fundamental. Therefore, in addition to the different grouping and the number of objects from all classes, the distance between these objects is also relevant for the quality of the detection system. The network must be able to correctly detect all objects, regardless of how close they are positioned to each other, and label them within the object bounding boxes.

For the industrial application of a detection system, some environmental factors play a major role in the detection of components. These are, among others, different lighting conditions, (different) shading, pollution, and damage of the components. In this work, all images were taken in a windowless room. An LED ring light was used as the light source to avoid varying light conditions. Another advantage is the reduction of different shadings, which would have to be encapsulated from detection anyway. Contamination of a component minimally affects the quality of the detection system. Since soiling only leads to a change in the color, but not in the geometry of a component, this feature difference seems to have a very small impact on the correct detection of a class and is thus irrelevant for object detection. The situation is different if a component is damaged. If the damage is too large, e.g., if there are long cracks or large notches on a component, this feature difference can be large because the geometry structure is very different compared to the undamaged component. This discrepancy can affect not only object recognition, but more importantly, the confidence value of the recognized object. If components with interfering factors are present, these can be used within the training data to extend the intuition of the detection system. By training the network for such interfering factors in advance, the risk of false detection within industrial use is avoided or minimized.

The confidence value should always be evaluated with care. An accurate confidence value should only be considered when the detected object is assigned to the correct class. An incorrect object detection, i.e. when a detected object is assigned to an incorrect class, may well have a very stable confidence value. This can occur, among other things, if there is not enough training data and the network selects wrong approaches to learn the objects, e.g., recognizing the objects based on the object bounding boxes. This is problematic because an object bounding box is not a feature of an object, but only the location of an object within a record.

*C. Performance*

The performance assesses the performance of the set-up system. The detection depends on some main factors, e.g., camera selection and computing capacity, as well as the extension of the system.

Extending the system, especially by adding new classes, is basically not a problem, since the system is already quite stable due to the different classes, which differ mainly by their asymmetric features such as geometry and contour. Thus, ambiguous object constellations are avoided for objects that differ slightly based on their characteristics. Of course, the newly added classes would have to go through the training process like all other classes before them and be signed off from all positions and views to provide the network with enough intuition to detect them. Using the open-source library OpenCV, the model can be modified and adapted to the task at hand at any time.

Camera selection is an important factor for detection. The current state of the art requires a certain resolution as a basic prerequisite for object detection. With a resolution of 1536 x 1536, the camera used for the training and test scenarios meets all requirements in terms of timeliness. With higher resolutions, more computing capacity is required. Therefore, this influence is stronger compared to the selection of the camera. The advantage of the used single-stage detector, the YOLOv4 algorithm, is the speed with which fast predictions can be made. This is especially evident in the developed application, as the computation time of the Raspberry Pi V4 for one image is less than one second. Thus, real-time detection is possible with the developed detection system. Therefore, the developed application is well suited for industrial application.

*D. Limits and Singularities of The Detection System*

The functionality of the model was intensively tested using the different test scenarios. Through the differently composed class selection within the second and third test scenarios, the model was critically examined and carefully tested. Already in the first test scenario, the network was able to recognize each class individually in different positions and views. The confidence value was 99 to 100 percent. In the next two scenarios, the number of classes to be detected was increased and the composition of the objects from the different classes was random, as was the distance between the objects. All objects were almost always detected correctly, i.e., assigned to the correct class. The confidence value of the detection was between 98 and 100 percent. Only at distances where several objects were very compact next to each other, false detections occurred in some cases, where the object bounding box did not turn out to be stable. Several factors may play a role in the false detections. The used Raspberry Pi with a Pi Camera V2 is controlled manually for zooming in. This can lead to distortions of the images, which is noticeable during object detection. Another factor is the image contrast and lighting conditions. These interfering factors can, as in the second and third application scenarios, lead to difficulties in recognizing components in some image evaluations, which results in the false detection of objects. The used Raspberry Pi with a Pi Camera V2 is controlled manually for zooming in. This can lead to distortions of the images, which is noticeable during object detection. Another factor is the image contrast and lighting conditions. These interfering factors can, as in the second and third application scenarios, lead to difficulties in recognizing components in some image evaluations, which results in the false detection of objects.

Two further points of criticism should be mentioned: First, the object bounding boxes for some views, e.g., the side view of some objects, are somewhat larger. Secondly, the rectangular object bounding boxes do not adapt to the orientation of the detected object.

It can be concluded that Deep Learning is suitable for automated object recognition of unsorted components. The prerequisite is a data set that has sufficient records of all objects to train the network. Based on the predefined test scenarios, it is possible to judge the quality of the detection system as well as the data set and the trained network.

VIII. SUMMARY

In this work, we investigated the suitability of a deep learning algorithm for automated object detection of different components from a subassembly. For this purpose, the single-stage detector YOLOv4 in the darknet environment was analyzed as a deep-learning model. The studied model turned out to be an accurate method for solving the detection of different components with different features. The selected one-step detection process supports correct and, most importantly, fast classification and generation of object bounding boxes for all objects. The created model meets all criteria for automated object detection of different components. The recordings of these components, individually or as a group, were created using the hardware, Raspberry Pi, and Pi Camera V2.

The detector behaves stably during automated upper detection. The quality and the computing time are not negatively influenced by the number of objects to be detected and are constantly very stable. This also applies to the positioning of the objects within the detection plane. The results do not deteriorate, regardless of the view or the positioning of the objects. Interfering factors like shading of the objects or a close composition of several or even all classes do not seem to exert a negative influence.

Different test scenarios were performed within a two-dimensional detection plane. In the first scenario, the trained network was tested for the detection of individual classes with different orientations. This test scenario represented a mapping of the training process, especially since the recording of all classes as individual objects was processed in all positions and views in the training process. All objects were assigned to the correct class with a confidence value of 100 percent, with a few exceptions of 99 percent, and all object bounding boxes encompassed all objects accurately. In the second test scenario, the complexity was increased significantly, and the developed network had to correctly recognize multiple classes from different positions and views with varying numbers of objects. The distances of the objects to each other, which were in part close, were all classes correctly recognized predominantly with a confidence value of 99 to 100 percent. Only for some objects did a slightly worse confidence value result for some positions and views. Likewise, all object bounding boxes turned out to be very stable. In the third test scenario, the complexity from the second test scenario was extended by having all 13 classes present in all records. Because all existing classes were present in different orientations and numbers of objects, among other things, a tight composition resulted and, in some places, objects were lined up. Nevertheless, all classes were recognized correctly almost throughout, and the confidence value was predominantly 98 to 100 percent. Like the second application scenario, some objects in some positions and views had a slightly worse confidence value. In a single case, this was 86 percent. All objects were correctly recorded and labeled by the object bounding box. Only on two recordings for two different objects, these were not detected by the object bounding box and thus not detected.

Based on the defined test series, a use for automated object recognition could be sufficiently proven for the developed deep learning model. Since the computing time for each image is less than one second, real-time recognition with the developed detection system is possible as an application for industrial use.

## IX. Conclusion

For the use case tested in this work, extensions can be defined to test the accuracy of the detection system for other scenarios from industrial applications.

An imaginable scenario is the presence of an object that is not present in any class, e.g., this could be a ballpoint pen that fell out of an employee's pocket onto the detection plane during assembly. The designed network would need to be able to detect this object with an object bounding box and label it as unknown. Another scenario would be to examine objects with damage. Some of these could already be used during training. Since all types of damage can occur on a component in industrial applications, the correct detection and the confidence value would have to be examined in more detail using different test scenarios.

For an extension in terms of accuracy, a detection system with image segmentation would be conceivable. By dispensing with object bounding boxes, less positioning of the objects would be necessary, since the image segmentation accurately detects the complete contour of an object pixel by pixel. For this purpose, a single-level detector, such as YOLOv4, which was used in this work, would be recommended because the image segmentation approach involves increased detection time. A single stage detector is characterized by its speed and the extent to which the increased detection time could be reduced would need to be investigated, making the image segmentation approach useful for real-time detection of objects within industrial applications.

For all extensions presented, an extension of the Convolutional Neural Network would also be conceivable. The extension from a two-dimensional to a three-dimensional level would be recommended for stacked objects. Different scenarios could be used to evaluate the correct recognition and confidence value of each stacked object.

Finally, due to the versatile implementation possibilities of a detection system, the expansion of the developed system for an augmented reality application for components of any kind would be conceivable. For such a research initiative, however, corresponding applications would have to be coordinated for the respective field of application.